\documentclass[aps,epsfig,floats,pre,twocolumn,showpacs]{revtex4-1}
\usepackage{graphicx}
\usepackage[latin2]{inputenc}
\usepackage{amsmath}
\usepackage{amssymb}
\usepackage{epsfig}
\usepackage{bm}
\usepackage{color}

\begin{document}
\title{Migration of Cytotoxic T Lymphocytes in 3D Collagen Matrices}

\author{Z. Sadjadi$^*$}
\author{R. Zhao}
\author{M. Hoth}
\author{B. Qu}
\author{H. Rieger}

%\address{Department of Theoretical Physics and Center for Biophysics, 
%Saarland University, D-66123 Saarbr\"ucken, Germany}
%\address{Center for Integrative Physiology and Molecular Medicine, School of Medicine, Saarland University, Homburg, Germany}

\begin{abstract}

%We investigate the migration of cytotoxic T lymphocytes in 3D collagen networks. 
 
%$Immune cells fullfil their functions by migrating in complex biological microinvironments. 
To fulfill their killing functions, cytotoxic T lymphocytes (CTLs) need to migrate to search for their target cells in complex biological microenvironments, a key component of which is extracellular matrix (ECM).
The 
mechanisms underlying CTL's navigation are not well understood so far. Here we use a collagen assay 
as a model for the ECM and analyze the migration trajectories of primary human CTLs in collagen 
matrices with different concentrations. We observe different migration patterns for individual 
T cells. Three different motility types can be distinguished: 
slow, fast and mixed motilities. Slow CTLs remain nearly stationary within the collagen matrix and show slightly anti-persistent motility, while 
the fast ones move quickly and persistent (i.e. with not too large turning angles). The dynamics of the mixed type consists of periods of slow and fast 
motions; both states are persistent, but they have different persistencies. The dynamics can be 
well described by a two-state persistent random walk model. We extract the parameters of the model 
by analyzing experimental data. The mean square displacements predicted by the model and those measured experimentally are in very good agreement, without any fitting parameter. Potential reasons for the observed two-state motility are discussed. T cells dig the collagen during their migration and form channels, which facilitate the movement of other CTLs in the collagen network.

\end{abstract}
\maketitle
%\begin{widetext}
%\noindent {\bf SIGNIFICANCE} \\
%Cytotoxic T lymphocytes (CTLs) are key players in the adaptive immune system to 
%eliminate tumor cells or pathogen-infected cells. They fulfill their functions by migrating in complex biological microinvironments in the body which mainly consists of Extracellular matrix (ECM). The mechanisms underlying their navigation and search strategy are not well understood so far. To better understand the migration of CTLs in ECM, we analyze their trajectories within 3D collagen networks. Our main observation is that CTLs tear the collagen fibers forming channels, which facilitate movement of other T cells in collagen network. We describe the dynamics by a two-state random walk model, which reproduces the behavior of CTls without any free parameters.
%\end{widetext}
%

\section{Introduction}
\label{sec:Introduction}

Cytotoxic T lymphocytes (CTLs) are fully activated CD8$^+$ T cells, which are key 
players in the adaptive immune system to eliminate tumorigenic or pathogen-infected 
cells \cite{Zhang2011}. CTLs need to find their cognate antigens or cancerous cells, which 
are low in number\cite{Krummel2016,Fearnlay1999}. Thus, the ability of CTLs to 
efficiently navigate and search is crucial for an efficient immune response. Migration 
behavior of immune cells in the body and the search strategies they might follow 
is currently of great interest in physics and biology\cite{Fricke2016, Baumgart2019, Moses2019}. Migration of naive T cells 
in lymph nodes reportedly follows a Brownian or even subdiffusive dynamics 
\cite{Bajenoff2006,Beauchemin2007,Preston2006}, but switchings between fast and slow  motility 
modes have been also observed \cite{Miller2002}. 
Outside the lymph node, activated T cells destined to find their targets in 
peripheral tissues, most of which are characterized by dense extracellular matrix 
(ECM) \cite{Crapo2011}. Here a faster migration, e.g.\ via longer phases of 
superdiffusive dynamics or less switchings to the slow diffusive mode, is 
advantageous to be able to scan a larger tissue efficiently. For instance, it 
was reported that the dynamics of CD8$^+$ T cells in infected brain tissue 
resembles a Levy walk \cite{Harris2012}.

The extracellular matrix (ECM) mainly consists of collagens and is the major 
component of all tissues and organs. It has essential regulatory roles in nearly 
all cellular functions. Some collagens have inhibitory effects on the function 
of different immune cells \cite{Kelli1990,Rygiel2011}. In various types of cancer, 
the collagen network becomes dense, stiff and linearized in the vicinity of 
tumors, facilitating the transport of cancerous cells and making the ECM an important 
player in cancer metastasis, intravasion and prognosis \cite{Xu2019,
Zhou2017,Han2016,Miyazaki2018,Provenzano2006,Pengfei2012}. Additionally, the 
activity of T cells is also influenced by the density of the collagen matrix in 
tumors \cite{Kuczek2018}. Recently, different immune cells have been investigated 
in immunotherapy studies as potential drug delivery vehicles into the tumors 
\cite{Eyileten2016,Xie2016}. Understanding the migration and interactions of immune 
cells in collagen networks is crucial to unravel the underlying details of 
the immune response and design effective treatment strategies.

Collagen-based assays have been used to investigate the migration of lymphocytes 
in ECM and study the possible underlying mechanisms of immune interactions with 
ECM \cite{Haston1982,Schor1983,Kelli1990,Friedl1998,Artym2010,Olofsson2019}. In 
a recent study, collagen hydrogels were employed to compare migration patterns 
of human CD8$^+$ T cells in aligned and nonaligned collagen fibers as  
microenvironments resembling tumor cells and normal tissues, respectively 
\cite{Hawley2019}.

In this study, we use bovine collagen to construct a 3D environment in vitro 
as a model for the ECM. The trajectories of primary human CTLs 
in collagen matrices with different concentrations are analyzed for two blood 
donors. We find three different types of motion in both donors: The migration 
of CTLs can be categorized into slow, fast and mixed sub-groups. In the latter 
case, T cells switch between two persistent states with different persistencies. 
Such two-state motilities may point toward the possible involvement of search 
optimization processes \cite{Benichou}. We employ a stochastic random walk model 
developed for active processes with two arbitrary states \cite{Shaebani19} to 
describe the trajectories of of T cells migrating in collagen. By extracting the model parameters from 
experimental data, we analytically reproduce the MSD results obtained from experiments, 
remarkably without any free parameter.

\begin{figure*} 
\centering
\includegraphics[scale=0.9,angle=0]{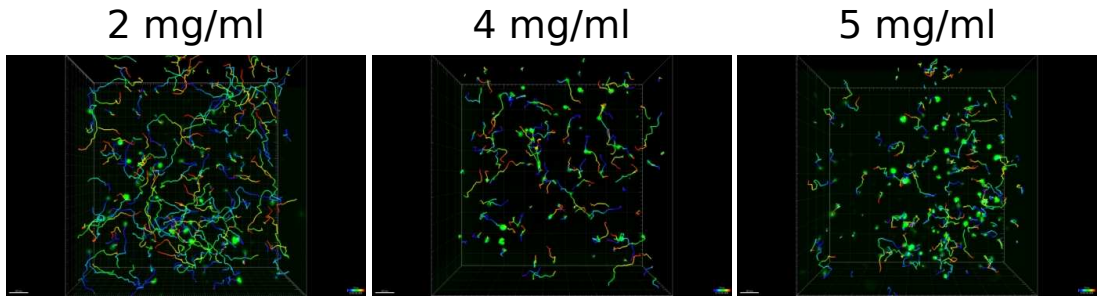}
\caption{
CTL migration trajectories measured by lightsheet microscopy in in three dimensional collagen matrices with different concentrations. 
The nuclei of human primary CTLs are labeled with overexpressed Histone 2B-GFP 
(green). CTL migration trajectories are tracked automatically using Imaris. 
Scale bars are 40 $\mu m$.}
\label{Fig:Exp1}
\end{figure*} 

\section{Materials and methods}
Ethical considerations:
Research carried out for this study with human material (leukocyte reduction system chambers from humanblood donors) is
authorized by the local ethic committee (decleration from 16.4.2015 (84/15; Prof. Dr. Rettig-St\"urmer)).

Human primary cytotoxic T lymphocytes (CTL) isolation, stimulation and nucleofection of CTLs: 
Peripheral blood mononuclear cells (PBMC) were obtained from healthy donors as previously 
described \cite{Kummerow14}. Human primary CTLs were negatively isolated from 
PBMC using Dynabeads$^{TM}$ Untouched$^{TM}$ Human CD8 T Cells Kit (ThermoFisher Scientific) or CD8$^+$ T Cell 
Isolation Kit, human (Miltenyi Biotec), stimulated with Dynabeads$^{TM}$ Human T-Activator CD3/CD28 (ThermoFisher 
Scientific) in AIMV medium (ThermoFisher Scientific) with 10\% FCS and 33 U/mL of recombinant 
human IL-2 (ThermoFisher Scientific). 48 hours after stimulation beads were 
removed and $5 \times 10^6$ CTLs were electroporated with 2 $\mu g$ plasmid (H2B-GFP to label nucleus 
\cite{Kanda98} or pMax-mCherry to label cell body) using the Human T cell 
nucleofector kit (Lonza). Medium was changed 6h after nucleofection and transfected CTLs 
were maintained in AIMV medium (ThermoFisher Scientific) with 10\% FCS and 33 U/mL of recombinant 
human IL-2 (ThermoFisher Scientific). Cells were used 24-36 hours after nucleofection \cite{Rouven17}.  
\subsection*{3D live cell imaging with lightsheet microscopy}
3D live cell imaging using lightsheet microscopy was conducted mainly as described previously \cite{Rouven18}. Briefly, human primary CTLs were resuspended first in PBS (ThermoFisher Scientific), afterwards neutralized 
collagen stock solution (bovine collagen type I, 8 mg/mL, Advanced Biomatrix) was added to a final concentration of 
2 mg/ml, 4 mg/ml, or 5 mg/ml collagen with a cell density of $10 \times 10^6$ cells/ml. 
20 $\mu l$ of this cell/collagen mixture was loaded in a capillary. The capillary was closed and incubated 
for 60 min in an incubator.    Afterwards, the polymerized collagen rod was pushed out hanging in the medium at $37C$ with 5\% CO$_2$ for equilibration for another 60 min.                
To visualize collagen structure,  analyzed collagen matrix was stained with 50 $\mu$g/ml Atto 488 NHS ester (ThermoFisher Scientific) 
in AIMV medium at room temperature after collagen polymerization. 
Afterwards, matrix was washed in AIMV medium. 
After collagen polymerization, cells in matrix with or without 
collagen staining were incubated in AIMV medium with 
10\% FCS at $37^{\circ}C$  with 5\% CO$_2$ for 1 hour.  Afterwards, 
the migration of cells was visualized by lightsheet 
micoscopy (20$\times$objective) at $37^{\circ}C$ with a z-step 
size of 1 $\mu m$ and time interval of 30 seconds. 
The CTLs were transiently transfected with histone 2B-GFP or mCherry. The 
migration trajectories were tracked and analyzed 
using Imaris 8.1.2 (containing Imaris, ImarisTrack, ImarisMeasurementPro, 
ImarisVantage; Bitplane AG, software available 
at http://bitplane.com) \cite{ Rouven18}.

\hspace{0.5cm}
\section{Experimental results}
\begin{figure} 
\centering
\includegraphics[scale=0.75,angle=0]{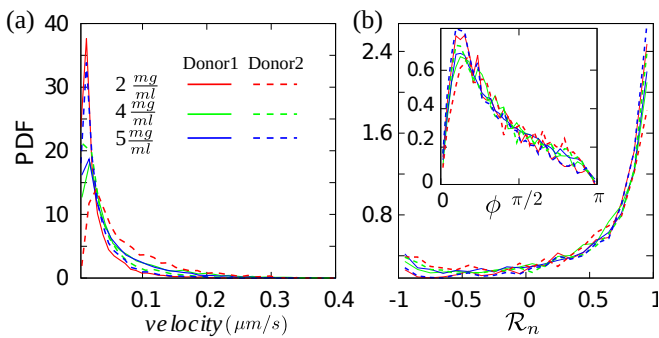}
\caption{Distributions of (a) velocity, and (b) persitency of T cells in collagen 
networks with different concentrations. Turning-angle distributions are shown in 
the inset.}
\label{Fig:Dists}
\end{figure}

To investigate migration patterns of CTLs in a 3D environment, we embedded primary human CTLs into collagen matrices and visualized their movements using lightsheet microscopy (Fig. \ref{Fig:Exp1}). 
Different concentrations of collagen mimic the ECM of normal tissue (2 mg/ml), soft solid tumor (4 mg/ml) and hard solid tumor (5 mg/ml), respectively\cite{Cox17,Ayyildiz15,Wang13}.
The experimental trajectories consist of a set of T cell 
positions recorded after equal time intervals. Every two successive recorded positions are used to calculate the 
instantaneous velocity, and every three of them to extract the corresponding 
turning angle $\phi$. The instantaneous persistency can be defined as $\mathcal{R}_n
{=}\cos\phi$. The global persistency $\mathcal{R}$ is the average over all $\mathcal{R}_n$s. 
The resulting parameters are summarized in Table~\ref{Tab:statistics}. 
The average velocity is higher at lower densities of collagen as expected, but 
the persistency of T cells is independent of the collagen concentration.
The cross correlation between velocity and persistency, $CC_{v,\mathcal{R}}=(\langle v \mathcal{R}\rangle- 
\langle v\rangle\langle \mathcal{R}\rangle)/\sigma_v \sigma_{\mathcal{R}}$ , which shows how these two 
parameters are related, also shows no systematic dependence on the collagen density. 
This value is in all cases positive, indicating that faster T cells move more persistent then slow ones. 
%The velocity auto-correlation function $\langle V(0){\cdot}V(t)\rangle$ as well. 
The distributions of velocity, turning-angle and persistency are shown in Fig.~\ref{Fig:Dists}. 
The distributions of $\mathcal{R}_n$ and $\phi$ show an overall persistent random 
walk for all data sets by a tendency to turn with an angle around 0.4 to 0.5 
radian (corresponding to a persistency around $\mathcal{R}{\simeq}0.9$). This angle 
could reflect the structure of the collagen network.

\begin{figure} 
\centering
\includegraphics[scale=0.8,angle=0]{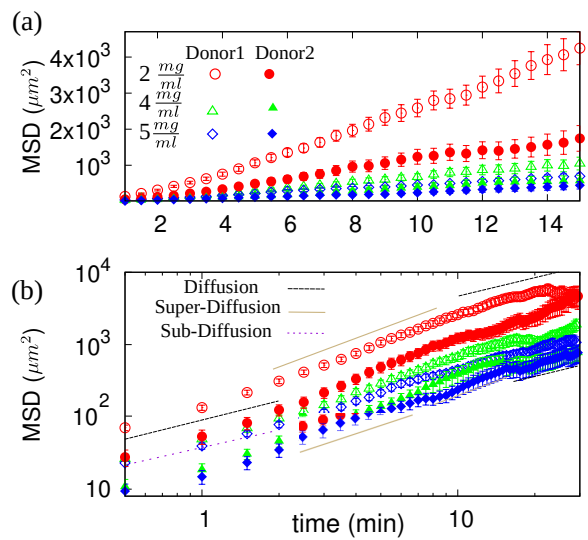}
\caption{Mean square displacement of CTLs in collagen matrices with different 
concentrations in (a) normal and (b) log-log scales.}
\label{Fig:MSD}
\end{figure}

\textit{Mean square displacement.}
To better understand the dynamics of T cells in different donors and collagen 
concentrations we analyze the mean square displacement (MSD) separately for 
each experimental condition. Figure~\ref{Fig:MSD} shows the MSD of two donors 
in different collagen concentrations. In all cases an initial diffusion or 
sub-diffusion is  followed by a slight supper-diffusive motion. Eventually a crossover 
to diffusion is observed which is expected because on longer time scales the 
trajectories are randomized and the orientational memory is lost.

   \begin{widetext}

\begin{table}
\caption{Key statistical parameters of  T cells in collagen matrices with 
different densities.}
\begin{tabular}{c|c c c||c c c}
 %&   Donor 1& &  &Donor 2 & & &
  &  & Donor 1 &  &  & Donor 2 &  \\ 
density $(\frac{m\text{g}}{\text{ml}})$  & 2  & 4 & 5 &  2 & 4 & 5 \\  
\hline 

velocity $(\frac{\mu\text{m}}{\text{s}})$   & 0.10$\pm$ 0.04  & 0.06$\pm$ 0.03 & 0.05$\pm$ 0.05 & 0.07$\pm$ 0.03 & 0.04$\pm$ 0.02 & 0.03$\pm$ 0.01   \\ 

persistency $\mathcal{R}$   & 0.30$\pm$ 0.30  & 0.36$\pm$ 0.29 & 0.35$\pm$ 0.29 & 
0.30$\pm$ 0.30 & 0.36$\pm$ 0.30 & 0.44$\pm$ 0.25  \\

$CC_{v,\mathcal{R}}$   & 0.64  & 0.35 & 0.35 & 0.36 & 0.29 & 0.31 

\label{Tab:statistics}
\end{tabular} 
\end{table}

\end{widetext}

\begin{figure} [b]
\centering
\includegraphics[scale=0.63,angle=0]{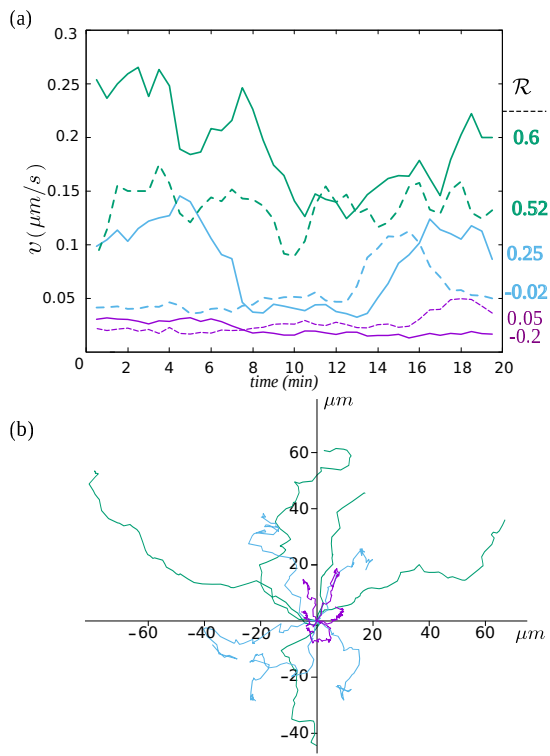}
\caption{(a) Typical velocities of CTL traces. Green, purple and blue colors 
correspond to fast, slow and mixed types of motility. (b) Typical trajectories 
of different motility types. The traces shown here are taken from  Donor 1 in 
collagen with a concentration of 2 mg/ml.}
\label{Fig:Types}
\end{figure}

\subsection{Three motility groups can be distinguished in CTL dynamics.}
\label{twostate}
Single track analysis of CTL trajectories reveals that there are three different 
types of CTL trajectories: (i) slow T cells which perform a sub-diffusive motion 
and their velocities always remain below a threshold value, (ii) a faster group 
with velocities always above a threshold value, and (iii) the third group with 
velocities which switch between fast and slow modes. Both donors have CTL of these 
three types, though with different proportions of them (shown in Table~\ref{Tab:percent}). 
The velocity evolution of typical tracks of T cells and a few trajectories for 
each cell motility type are shown in Fig.~\ref{Fig:Types}. 

\begin{figure}
\centering
\includegraphics[scale=0.7,angle=0]{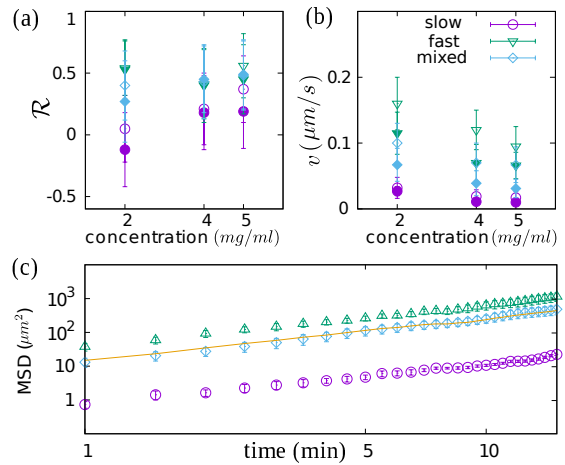}
\caption{(a) $\mathcal{R}$ and (b) velocity of different motility types in two 
donors represented by open and full symbols. (c) Mean square displacement of 
different motility types of donor 2 in collagen concentration 5 mg/ml. The 
solid line represents the MSD of all T cells.}
\label{Fig:Subtypes}
\end{figure}

We consider two velocity thresholds to distinguish the three types. All T cells 
whose maximum (minimum) velocity is less (more) than $v_{c1}$ ($v_{c2}$) are 
categorized as slow (fast) type. Trajectories with $v_{max}>v_{c2}$ and 
$v_{min}<v_{c1}$ lie in the mixed category. Around 6 to 8 percent of T cells 
belong to non of these groups with varying parameters $v_{c1}$ and $v_{c2}$. 
Figures~\ref{Fig:Subtypes}(a,b) summarize the 
average velocities and persistencies of different types in different collagen 
concentrations for both donors. A slight overal increase (decrease) in the 
persistency (velocity) is observed with increasing collagen concentration.

The mean square displacements of different types of motion are clearly distinguishable 
(see e.g.\ Fig.~\ref{Fig:Subtypes}(c)), indicating underlying differences between 
the dynamics of the three categories of motility.

\subsection{The two-state motility type}

Intermittent motion is widely observed in nature and it is shown that transitions 
between different internal states can help to optimize the search time \cite{Benichou}. 
In the following we study the mixed-velocity trajectories of 
T cells in more details.

\textit{ Mean square displacement.} The MSD of the mixed category coincides 
roughly with the total MSD, nearly in all cases (see the solid line in 
Fig.~\ref{Fig:Subtypes}(c) for one sample). This indicates that the frequency 
of fast or slow periods in the mixed case have similar statistics as the purely 
fast or slow types of motion. In the next section, we compare the MSD of the mixed trajectories 
with the prediction of a two state random velocity model.

\begin{figure} [h]
\centering
\includegraphics[scale=0.65,angle=0]{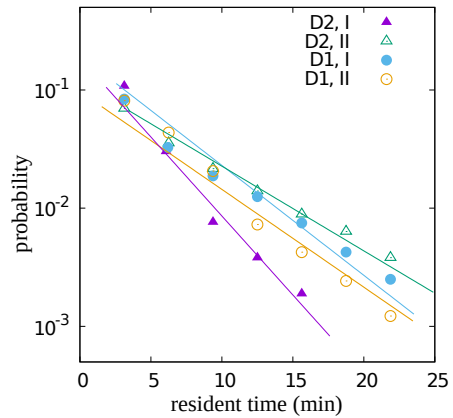}
\caption{Sojourn time distributions in the two states of mixed type of T 
cells in collagen with 4 mg/ml concentration. The lines are the corresponding 
theoretical estimate for each case (same color). D stands for donor.
}
\label{Fig:Tdist}
\end{figure}

\textit{Exponential distribution of sojourn time in each state.}
The distribution of the times that the T cells remain in one state before they switch, the so-called sojourn times, follow an 
exponential decay as shown in figure~\ref{Fig:Tdist}. In this example the sojourn time distribution 
of T cells in  the different states of the mixed T cell migration type of two donors 
in 4 mg/ml collagen concentration is plotted. The exponential decay of these distributions 
indicates that the transition probabilities are time-independent. This allows us to model the 
motion of T cells which switch to another motility state with a constant transition probability. 
%Figure~\ref{Fig:Tdist} shows the sojourn time distribution 
%of T cells in  the different states of the mixed T cell migration type of two donors 
%in 4 mg/ml collagen concentration. 
\begin{table} 
\caption{Percentage of different motility types of CTLs in collagen matrices 
with different densities.}
\begin{tabular}{c|c c c||c c c}
&  & Donor 1 &  &  & Donor 2 &  \\ 
 density $(\frac{m\text{g}}{\text{ml}})$  & 2  & 4 & 5 &  2 & 4 & 5  \\ 
\hline 

slow \%   & 33 & 43 & 48 & 15 & 27 & 26  \\ 

fast \%   & 35 & 14 & 16 & 30 & 25 & 24  \\

mixed \%   & 26 & 37  & 25 & 48 & 40 & 43

\label{Tab:percent}
\end{tabular} 
\end{table}
\section{Two-state persistent random walk model}

In the following we show that the experimentally measured T cell trajectories are well described by a stochastic process that involves a persistent random walk with two different motility states \cite{Shaebani19}.
We confine ourselves to a two-dimensional model to derive an analytical formula for the mean square displacement of one Cartesian coordinate $\langle x^2(t) \rangle$, which then multiplied by 3 to give the prediction for the MSD in three dimensions $\langle r^2(t)\rangle = 3\langle x^2(t) \rangle$\cite{Sadjadi15}. 

We adopt a discrete-time approach, since it enables us to reproduce the detailed particle 
dynamics obtained from analyzing the experimental data. As the trajectories in experiments 
comprise a regularly recorded set of particle positions, a continuous-time description fails 
to capture the short time behavior. 
%For example, let us consider a Brownian motion, simply 
%described by an ordinary diffusion in continuous time. 
By tracking the particle with a very 
fast camera, an extremely fine time resolution $\Delta t$ compared to the 
characteristic time of particle wiggling can be obtained. In such a case, the consecutive 
orientations are correlated and the motion is supperdiffusive at short times with a crossover 
to normal diffusion at longer times. A discrete time formalism is able to capture such transient dynamics. 
Note that the initial dynamics in the limit $\Delta t{\rightarrow}0$ is even a purely ballistic 
motion, thus, the discrete-time approach in this limit is inequivalent to the continuous-time description. 

We consider persistent motions characterized by velocity distributions 
$F\!\!_{_{_{\text{I}}}}\!(v)$ and $F\!\!_{_{_{\text{I\!I}}}}\!(v)$, and turning-angle 
distributions $R_{_{_{\text{I}}}}\!(\phi)$ and $R_{_{_{\text{I\!I}}}}\!(\phi)$ for states ${\text{I}}$ 
and ${\text{I\!I}}$. We introduce the transition probabilities 
$\kappa_{_{_{\text{I}\rightarrow\text{I\!I}}}}$ and $\kappa_{_{_{\text{I\!I}\rightarrow\text{I}}}}$ 
for switching from state $\text{I}$ to $\text{I\!I}$ and vice versa. These probabilities are 
estimated by the inverse of sojourn time in the two states of mixed trajectories, e.g.
$\kappa_{_{_{\text{I}\rightarrow\text{I\!I}}}}\!\!\sim\!\langle\tau\rangle_{\text{I}}^{-1}$.
Constant probability transitions 
$\kappa_{\text{I}\rightarrow \text{II}}$ ($\kappa_{\text{I\!I}\rightarrow 
\text{I}}$) lead to an exponential distribution of the sojourn time 
$\mathcal{F}\!_{_{\text{I}}}{(\tau)}{\sim}\,\text{e}^{\ln(1{-}\kappa_{\text{I} 
\rightarrow \text{I\!I}}\!)\;\!\tau}$ ($\mathcal{F}\!\!_{_{\text{I\!I}}}{(\tau)}{\sim}\,
\text{e}^{\ln(1{-}\kappa_{\text{I\!I}\rightarrow \text{I}}\!)\;\!\tau}$). 
As a first approximation, $\mathcal{F}\!_{_{\text{I}}}{(\tau)}{\sim}e^{-\tau/\langle\tau\rangle_I}$ and 
$\mathcal{F}\!_{_{\text{I\!I}}}{(\tau)}{\sim}e^{-\tau/\langle
\tau\rangle_{I\!I}}$ are plotted in Fig.~\ref{Fig:Tdist}, which show a very 
good agreement with the  experimental resident time distribution in each state 
of the mixed motion.
Introducing the probability density functions $P_{t}^{\text{I}}(x,y|
\theta)$ and $P_{t}^{\text{I\!I}}(x,y|\theta)$ for the probability 
to find the walker at position $(x,y)$ along the direction $\theta$ 
at time $t$ in each of the motility states, the temporal evolution 
of the stochastic process can be described by the following set of 
coupled master equations:
\begin{eqnarray}
\begin{array}{ll}
P_{\!_{t{+}\Delta t}}^{\text{I}}(x,y|\theta) = \vspace{1mm} 
(1{-}\kappa_{_{_{\text{I}\rightarrow\text{I\!I}}}}) 
\!\! \displaystyle\int \!\!\! dv F\!\!_{_{_{\text{I}}}}\!(v) \!\!
\displaystyle\int_{\!{-}\pi}^{\pi} \!\!\!\!\!\! d\gamma \, R_{_{_{\text{I}}}}
\!(\theta{-}\gamma) P_{t}^{\text{I}}(x',y'|\gamma) \vspace{1mm} \\
\hspace{2cm}+ \kappa_{_{_{\text{I\!I}\rightarrow\text{I}}}} \!\! 
\displaystyle\int \!\!\! dv F\!\!_{_{_{\text{I}}}}\!(v) \!\!
\displaystyle\int_{\!{-}\pi}^{\pi} 
\!\!\!\!\!\! d\gamma \, R_{_{_{\text{I\!I}}}}\!(\theta{-}\gamma) 
\, P_{t}^{\text{I\!I}}(x',y'|\gamma),\\
P_{\!_{t{+}\Delta t}}^{\text{I\!I}}(x,y|\theta) = \vspace{1mm}  
(1{-}\kappa_{_{_{\text{I\!I}\rightarrow\text{I}}}}) \!\!
\displaystyle\int \!\!\! dv F\!\!_{_{_{\text{I\!I}}}}\!(v) \!\!
\displaystyle\int_{\!{-}\pi}^{\pi} \!\!\!\!\!\! d\gamma \, R_{_{_{\text{I\!I}}}}
\!(\theta{-}\gamma) P_{t}^{\text{I\!I}}(x',y'|\gamma) \vspace{1mm} \\
\hspace{2cm}+ \kappa_{_{_{\text{I}\rightarrow\text{I\!I}}}} \!\!
\displaystyle\int \!\!\! dv F\!\!_{_{_{\text{I\!I}}}}\!(v) \!\!
\displaystyle\int_{\!{-}\pi}^{\pi} 
\!\!\!\!\!\! d\gamma \, R_{_{_{\text{I}}}}\!(\theta{-}\gamma) 
\, P_{t}^{\text{I}}(x',y'|\gamma),
\end{array}
%\right.
\label{Eq:MasterEqs}
\end{eqnarray}
with $x'=x{-}v \Delta t\cos\theta$ and $y'=y{-} v \Delta t\sin\theta$. 
By solving these sets of master equations, one can evaluate arbitrary moments of the position of the walker, such as the mean square displacement.
Several mathematical techniques  like Fourier and $z$-transformation are used in order to solve equations \ref{Eq:MasterEqs}. The $z$-transformation $A(z)$ of an arbitrary function $A_n$ of a discrete variable $n=0,1,2,...$ is defined as  $A(z)=\sum\limits_{n=0}^{\infty}A_n z^{-n}$, which is equivalent to Laplace transformation in a continuous-time description. 
The exact result for the mean 
square displacement is obtained via inverse $z$ transformation of \cite{Shaebani19}: 
\begin{widetext} 
\begin{equation}
\begin{aligned}
&\frac{1}{(\Delta t)^2}\sum_{t=0}^{\infty} \!\! z^{-t} \! \langle x^2 \rangle(t) {=} \\
& \Big[\frac{z^2 \kappa_{_{_{\text{I\!I}{\!\rightarrow\!}\text{I}}}}}{G_0(z)}  +\frac{z(1-\kappa_{_{\text{I\!I}{\!\rightarrow\!}\text{I}}}-\kappa_{_{_{\text{I}{\!\rightarrow\!}\text{I\!I}}}})P_0^{\text{I}} }{z-1+\kappa_{_{_{\text{I\!I}{\!\rightarrow\!}\text{I}}}}+\kappa_{_{_{\text{I}{\!\rightarrow\!}\text{I\!I}}}}}  \Big] \times 
\Bigg[ \frac{z\left[z{-}(1{-}\kappa_{_{_{\text{I\!I}{\!\rightarrow\!}\text{I}}}})
\mathcal{R}_{_{_{\text{I\!I}}}}\right]}{(z{-}1)G_1(z)} \langle v \rangle^2_{_\text{I}} {+} \frac{z}{(z{-}1)G_1(z)} 
\kappa_{_{_{\text{I}{\!\rightarrow\!}\text{I\!I}}}} \mathcal{R}_{_\text{I\!I}} 
\langle v \rangle_{_\text{I}} \langle v \rangle_{_\text{I\!I}} 
{-} \frac{1}{z{-}1} \langle v \rangle^2_{_\text{I}} {+} \frac{1}{2(z{-}1)} \langle v^2 \rangle_{_\text{I}} \Bigg] \\
&{+} \Big[ \frac{z^2 \kappa_{_{_{\text{I}{\!\rightarrow\!}\text{I\!I}}}}}{G_0(z)}  +\frac{z(1-\kappa_{_{\text{I\!I}{\!\rightarrow\!}\text{I}}}-\kappa_{_{_{\text{I}{\!\rightarrow\!}\text{I\!I}}}})P_0^{\text{I\!I}} }{z-1+\kappa_{_{_{\text{I\!I}{\!\rightarrow\!}\text{I}}}}+\kappa_{_{_{\text{I}{\!\rightarrow\!}\text{I\!I}}}}}       \Big] \times 
\Bigg[ \frac{z\left[z{-}(1{-}\kappa_{_{_{\text{I}{\!\rightarrow\!}\text{I\!I}}}})
\mathcal{R}_{_{_{\text{I}}}}\right]}{(z{-}1)G_1(z)} \langle v \rangle^2_{_\text{I\!I}} {+} \frac{z}{(z{-}1)G_1(z)} 
\kappa_{_{_{\text{I\!I}{\!\rightarrow\!}\text{I}}}} \mathcal{R}_{_{\text{I}}} 
\langle v \rangle_{_\text{I\!I}} \langle v \rangle_{_\text{I}} 
{-} \frac{1}{z{-}1} \langle v \rangle^2_{_\text{I\!I}} {+} \frac{1}{2(z{-}1)} \langle v^2 \rangle_{_\text{I\!I}} \Bigg],
\end{aligned}
\label{Eq:MSD}
\end{equation}
\end{widetext} 
with
\begin{equation}
G_0(z)=(z-1)(z-1+\kappa_{_{_{\text{I\!I}{\!\rightarrow\!}\text{I}}}}+\kappa_{_{_{\text{I}{\!\rightarrow\!}\text{I\!I}}}}) \nonumber
\end{equation}
and
\begin{equation}
G_1\!(z)\!=\!\!\left[z{-}(1{-}\kappa_{_{_{\text{I\!I}{\!\rightarrow\!}\text{I}}}}) 
\mathcal{R}_{_{_{\text{I\!I}}}} \right]\!\!
\left[z{-}(1{-}\kappa_{_{_{\text{I}{\!\rightarrow\!}\text{I\!I}}}})
\mathcal{R}_{_{_{\text{I}}}} \right]\!-
\kappa_{_{_{\text{I}{\!\rightarrow\!}\text{I\!I}}}} \!\kappa_{_{_{\text{I\!I}{\!\rightarrow\!}
\text{I}}}} \!\mathcal{R}_{\text{I\!I}} 
\!\mathcal{R}_{\text{I}}\nonumber
\end{equation}

\begin{table} 
\caption{Parameters extracted from the experimental data of mixed type of  motility and used in the two-state model.}
\begin{tabular}{c|c c c||c c c}
&  & Donor 1 &  &  & Donor 2 &  \\ 
 density $(\frac{m\text{g}}{\text{ml}})$  & 2  & 4 & 5 &  2 & 4 & 5  \\ 
\hline 

$V_I (\mu m/s)$    & 0.066 & 0.047 & 0.048 & 0.047 & 0.023 & 0.020  \\ 

$V_{I\!I} (\mu m/s)$     & 0.138 & 0.120 & 0.111 & 0.099 & 0.061 & 0.055  \\

$V^2_I (\mu m^2/s^2)$   & 0.0189 & 0.0038 & 0.0037 & 0.0149 & 0.0009 & 0.0006  \\ 

$V^2_{I\!I} (\mu m^2/s^2)$     & 0.0510& 0.0198 & 0.0165 & 0.038 & 0.0055 & 0.0040  \\

$R_I$   & 0.15 & 0.35 & 0.38 & 0.19 & 0.26 & 0.35  \\ 

$R_{I\!I}$    & 0.62 & 0.59 & 0.59 & 0.33 & 0.56 & 0.69 \\ 

$\Delta t$ (min)    & 0.5 & 0.5 & 0.5 & 0.5 & 0.5 & 0.5  \\

$\kappa_{_{_{\text{I}{\!\rightarrow\!}\text{I\!I}}}}$ &0.20& 0.08 & 0.09 & 0.11 & 0.10 & 0.07  \\ 

$\kappa_{_{_{\text{I\!I}{\!\rightarrow\!}\text{I}}}}$ & 0.06  & 0.14 & 0.17 & 0.14 & 0.11 & 0.13 
\label{Tab:modelpara}
\end{tabular} 
\end{table}
 
\begin{figure}
\centering
\includegraphics[scale=0.8,angle=0]{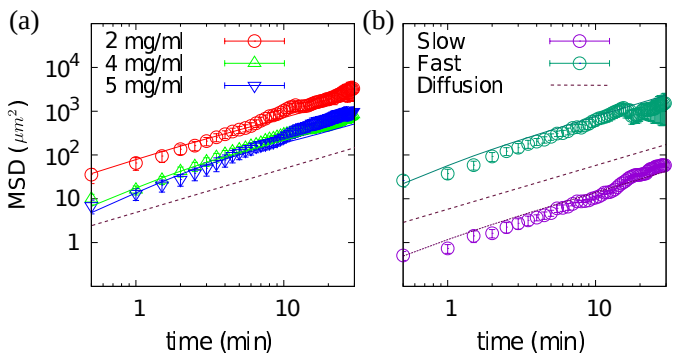}
\caption{ (a)  The mean square  displacement of mixed motility type in various collagen 
concentrations obtained from the experiments (symbols) and analytical approach (Equation 
\ref{Eq:MSD} with parameters extracted from experimental data summarized in table \ref{Tab:modelpara})  
(solid lines) for donor 2.
(b) The mean square  displacement of slow and fast motility types in collagen 
concentration 4 mg/ml. The solid lines represent the theoretical estimate of 
equation\ref{Eq:MSDreduced} withe parameters extracted from experiment (\ref{Tab:modelpara1State}). }
\label{Fig:MSDcompare}
\end{figure}

In equations \ref{Eq:MSD}, $\langle v \rangle_{\text{I}}$, $\langle v \rangle_{\text{I\!I}}$, 
$\langle v^2 \rangle_{\text{I}}$ and $\langle v^2 \rangle_{\text{I\!I}}$ are first and second 
moments of velocity of T cells in states $\text{I}$  and $\text{I\!I}$. $P_0^{\text{I}}=1-P_0^{\text{I\!I}}$ 
is the initial condition and shows the probability of starting the motion in state $\text{I}$ 
or equivalently  the percentage of all T cells in state $\text{I}$ at the beginning of tracking.  
$\mathcal{R}_{\text{I}}$ and  $\mathcal{R}_{\text{I\!I}}$ are the Fourier transform of distributions 
of turning angle $R_I(\phi)$ and $R_{I\!I}(\phi)$ in Eqs.\ref{Eq:MasterEqs}: 
\begin{equation}
\mathcal{R}_j=\!\int_{-{\pi}}^{{\pi}}\!\text{d}\phi\,e^{i\phi}R_{\text{j}}(\phi)=\langle cos\phi\rangle_j,  j\in \{\text{I},\text{I\!I}\}
\label{Rfourier}
\end{equation}

\begin{table} 
\caption{Parameters extracted from the experimental data of fast (up) and slow(down) types of motility.}
\begin{tabular}{c|c c c||c c c}
&  & Donor 1 &  &  & Donor 2 &  \\ 
 density $(\frac{m\text{g}}{\text{ml}})$  & 2  & 4 & 5 &  2 & 4 & 5  \\ 
\hline 

$V_I (\mu m/s)$    & 0.17 & 0.12 & 0.09 & 0.11 & 0.07 & 0.07  \\

$V^2_I (\mu m^2/s^2)$   & 0.033 & 0.019 & 0.013 & 0.018 & 0.008 & 0.006  \\

$R_I$   & 0.54 & 0.41 & 0.45 & 0.53 & 0.40 & 0.56  \\

$\Delta t$ (min)    & 0.5 & 0.5 & 0.5 & 0.5 & 0.5 & 0.5 \\
\hline 
\hline 

$V_I (\mu m/s)$    & 0.032 & 0.019 & 0.017 & 0.027 & 0.011 & 0.010  \\

$V^2_I (\mu m^2/s^2)$   & 0.0014 & 0.0006 & 0.0005 & 0.0010 & 0.0002 & 0.0001  \\

$R_I$   & -0.12 & 0.18 & 0.19 & 0.05 & 0.21 & 0.37  \\ 

$\Delta t$ (min)    & 0.5 & 0.5 & 0.5 & 0.5 & 0.5 & 0.5 
\label{Tab:modelpara1State}
\end{tabular} 
\end{table}

In the case of fast and slow types of motion, the master equations \ref{Eq:MasterEqs}, shrinks to:
\begin{eqnarray}
P_{\!_{t{+}\Delta t}}(x,y|\theta) = \vspace{1mm} 
\!\! \displaystyle\int \!\!\! dv F(v) \!\!
\displaystyle\int_{\!{-}\pi}^{\pi} \!\!\!\!\!\! d\gamma \, R
(\theta{-}\gamma) P_{t}(x',y'|\gamma) \vspace{1mm}.
\label{Eq:Masterreduced}
\end{eqnarray}
Where $P_{\!_{t}}(x,y|\theta)$ is the probability density of a T cell to arrive at position 
$(x,y)$ with direction $\theta$ at time $t$ and $F(v)$ and $R(\theta)$ are the distribution functions of velocity and turning angle, respectively.
The resulting MSD in this case will be: 
\begin{eqnarray}
\langle x^2\rangle(t)&=&(\frac{1}{2}\langle v^2\rangle+\frac{\mathcal{R}}{1-\mathcal{R}}\langle v\rangle^2)\Delta\!T t\\
&+&\frac{\mathcal{R}}{(1-\mathcal{R})^2}\langle v\rangle^2\Delta\!T^2(\mathcal{R}^{t/\Delta\!T}-1) \nonumber
\label{Eq:MSDreduced}
\end{eqnarray}

We do not use any fitting parameter, instead we extracted the model 
parameters from the experimental data analysis. Moments of velocity 
($\langle v \rangle_{\text{I}}$, $\langle v \rangle_{\text{I\!I}}$, 
$\langle v^2 \rangle_{\text{I}}$ and $\langle v^2 \rangle_{\text{I\!I}}$) in 
each state is calculated by averaging over local velocities of trajectories.
The persistencies $\mathcal{R}_{\text{I}}$ and  $\mathcal{R}_{\text{I\!I}}$ 
are measured by averaging over $cosine$s of turning angles (see Eq.\ref{Rfourier}).
The transition probabilities 
$\kappa_{_{_{\text{I}\rightarrow\text{I\!I}}}}$ and $\kappa_{_{_{
\text{I\!I}\rightarrow\text{I}}}}$ are the inverse of average sojourn time in 
states $\text{I}$ and $\text{I\!I}$, respectively. The initial condition $P_0^{\text{I}}$, 
only affects the short-time behavior of motion, We set this parameter to $0$ for all cases. 
This means that we assume all T cells start their motion in the faster mode.
All extracted model parameters are summarized in Tables \ref{Tab:modelpara} and 
\ref{Tab:modelpara1State} for mixed state and fast/slow states, respectively.

The time evolution of mean square 
displacement is shown in Fig.~\ref{Fig:MSDcompare} for donor 2.
Fig.~\ref{Fig:MSDcompare} shows exemplary match of the model with the data. 
We emphasize that no fitting has been made to capture this  agreement.

\begin{figure} [b]
\includegraphics[scale=0.7,angle=0]{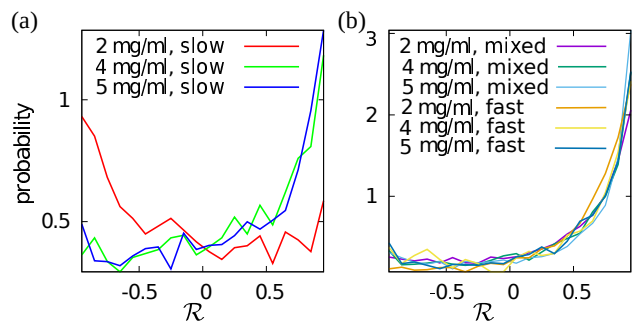}
\caption{ Probability distributions of persistency $\mathcal{R}$ for 
different motility types in various concentrations of collagen, for (a) 
slow and (b) fast and mixed types.}
\label{Fig:compareRdistTypes}
\end{figure}

\begin{figure*} 
\centering
\includegraphics[scale=0.9,angle=0]{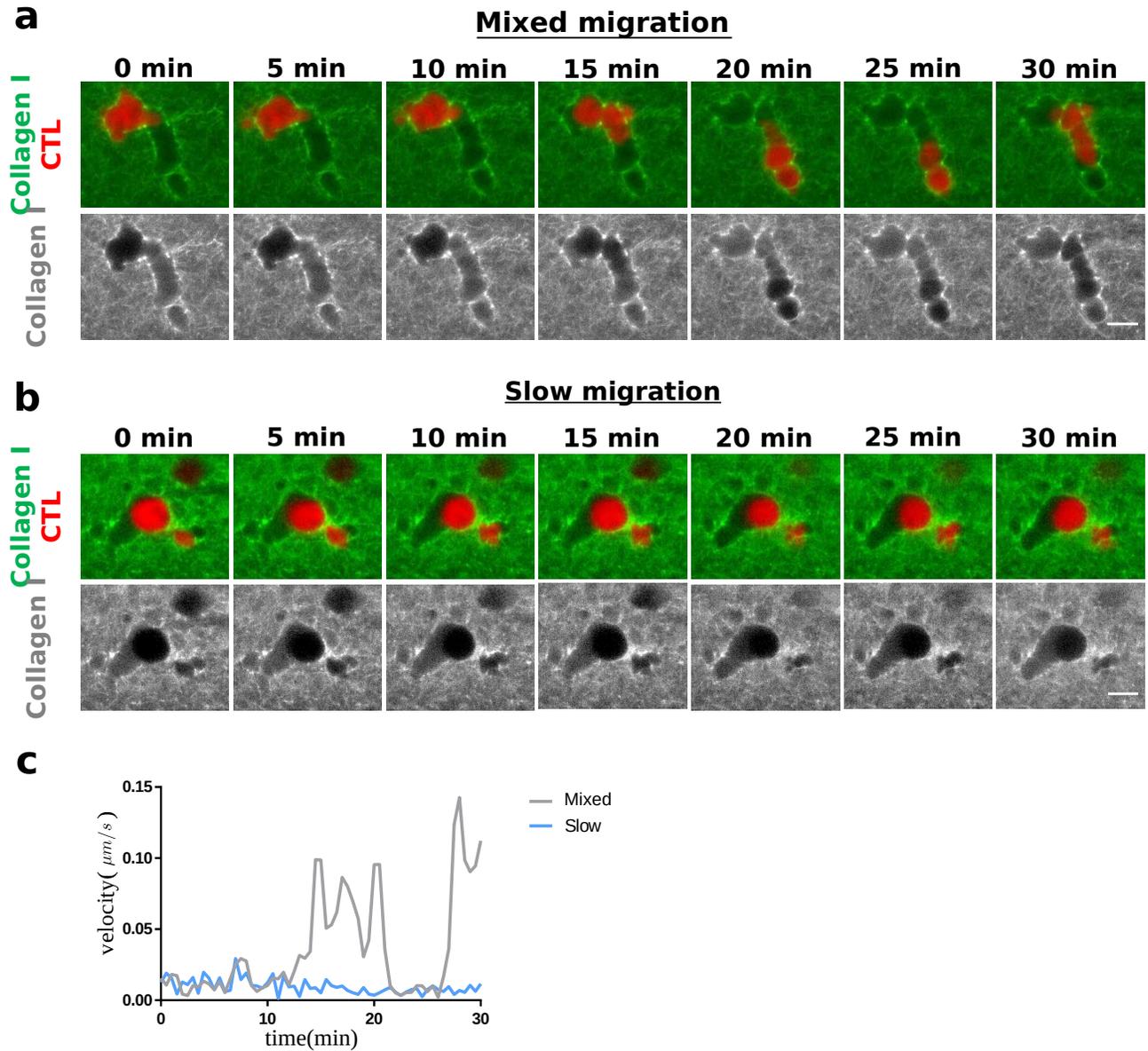}
\caption{Visualization of CTL migration in a collagen matrix. Primary Human CTLs were transiently
transfected with mCherry (red). Collagen (5 mg/ml) was stained with Atto 488 NHS ester (green or gray
as indicated). CTL migration was visualized at $37^{\circ}$C using lightsheet microscopy. Exemplary cells for mixed
and slow migration are shown in a and b, respectively (Movies in Suppl. info). One layer of z-stack is presented. c, Quantification
of migration velocity at all time points examined. CTL migration trajectories were tracked and analyzed
by Imaris. Scale bars are 10 $\mu m$.
}
\label{Fig:Timelaps}
\end{figure*} 

\section{Discussion}
We analyzed the trajectories of CTLs of two donors within 3D collagen 
matrices with different concentrations. We found three motility types in all experiments: slow, fast and mixed.  Similar motility types have been reported for natural killer (NK) cells in hydrogel collagen 
with a concentration of 3 $mg/ml$ in the presence of target cells \cite{Olofsson2019}. 
The similarity of the characteristics of CTL and NK cell trajectories 
points towards a common mechanism for migration of both cell types through collagen networks.

A plausible scenario is 
that the cells which arrive first in the collagen network perform a persistent 
random walk unless they move into denser areas of the network, where they 
become slower, but eventually find a way to move again, which leads to two-state motility. When the cells move through the collagen network, they leave
a channel by displacing or stretching collagen fibers. These channels facilitate the 
movement of other T cells, such that cells entering already existing channels move faster and tend to remain in the existing channel network.
They do not switch to slow movement and thus establish the fast type. The slow cells 
mainly remain in one part of the network and only ``wiggle" around and 
seem to be nearly immobile.

By studying the motion in different collagen densities, we found that the collagen concentration 
dependence of the persistency of slow T cells is different from the other types.
Figure~\ref{Fig:compareRdistTypes} shows the probability distribution of persistencies of 
the three different cell types. While the distributions for fast and mixed types show a persistent 
motion in all concentrations, slow T cells perform anti-persistent motion in 2 mg/ml 
collagen and become persistent in denser ones. 
We propose three explanations for these observations: 1) The slow cells have an activity level that 
is below that of the mixed and fast type, for instance due to an incomplete activation. 2) The slow 
T cells entered accidentally a region of the collagen matrix where still no channels produced by other 
T cells (within the 2 hours initiation time before measurements) and can only move forward by creating a 
new channel by deforming and/or destroying the collagen fibers, which is slow process 
as compared to fast migration with a channel. 3) The fast/mixed T cells and the slow T cells establish 
two different phenotypes. To test Explanation number 2, we fluorescently stained collagen and visualized 
the movements of CTLs using lightsheet microscopy. We found that during migration, CTLs could enter 
channels in collagen matrix. Inside the channel they had high speed and when leaving the channel 
significantly slowed them down (Fig. \ref{Fig:Timelaps}a, \ref{Fig:Timelaps}c (Suppl. Movie 1)). Slow CTLs appeared to 
be trapped in some channels and the corresponding speed stayed slow (Fig. \ref{Fig:Timelaps}b, 
\ref{Fig:Timelaps}c (Suppl. Movie 2)). These results support the Explanation 2. Still, Explanations 1 and 3 should not 
be excluded. The morphology of the "channel" visible in Fig. \ref{Fig:Timelaps} is incompatible with a randomly generated filament network (see background). The latter has of course randomly distributed regions with  higher and lower filament density but elongated cylindrical tunnels as visible in Fig. \ref{Fig:Timelaps}  with a diameter of approximately equal to the diameter of a T cell and completely 
void of filaments cannot occur with significant probability by chance. These cylindrical
tunnels are also unlikely to be produced by deformation: collagen fibers
are elastic and will at least partially spring back once T cell has passed.
More plausible appears to us the hypothesis that these channels have been
produced by either T cells degrading the local matrix by secretion of matrix 
metalloproteases or by T cells tearing matrix apart by exertion of mechanical forces during the 2 hours before the observation and
tracking was started - and leaving behind elongated, cylindrical tunnels
of approximately the same diameter as T cells. Concerning the former option, it is reported 
that treatment of MMP inhibitor in human CD4 T-blast does not affect T cell speed \cite{Wolf13}(Fig. 5D). 
In CD4$^{+}$ T cells, MMP2 and MMP9 is expressed \cite{Edsparr11} , which do not degrade collagen type I 
\cite{Agata16}, the type we used in our model. The collagen type I-degrading MMPs (MMP1, MMP8, MMP12 and MMP14) 
are not expressed in bead-stimulated primary human CD8$^+$ T cells \cite{Eva}. Due to lack of MMPs, channel formation most probably proceeds via collagen filament deformation or destruction rather than degradation. 
 It would be rewarding to catch a T cell in the act of deforming the local matrix
and analyse it with time lapse microscopy - but we have to leave this endeavor 
for future experiments.  \\
%We have to leave the clarification to future experiments that are able to correlate the paths of the T cells with existing channels in the three-dimensional network of collagen fibers.

%
%
%\noindent {\bf SUPPLEMENTARY INFORMATION} \\
%Supplementary information accompanies this paper, including two movies. \\
%
%\noindent {\bf COMPETING FINANCIAL INTERESTS} \\
%The authors declare no competing financial interests. \\
%
%\noindent {\bf AUTHOR CONTRIBUTIONS}\\
%\noindent Z.S., B.Q., M.H.\ and H.R.\ designed the research. 
%R.Z.\ and B.Q.\ performed the experiments. Z.S.\ 
%analyzed the experimental data and employed the analytical model. All authors contributed to the interpretation 
%of the results. Z.S., B.Q.\ and H.R.\ wrote the manuscript. Correspondence should be addressed to Z.S. (sadjadi@lusi.uni-sb.de).\\

\noindent {\bf ACKNOWLEDGEMENTS}\\
\noindent We acknowledge financial support from Collaborative Research Center SFB 1027, M.H. from BMBF grant 031L0133 and R.Z. from HOMFOR2018 grant.

\end{document}